\documentclass[10pt, conference, compsocconf]{IEEEtran}
\usepackage{amsmath}
\usepackage{booktabs}
\usepackage{bm}

\ifCLASSINFOpdf
\else
\fi
\usepackage{graphicx} 
\usepackage{subfigure}

\usepackage{algpseudocode}
\usepackage{amsmath}

\usepackage{algorithm}
\usepackage{multirow}
\usepackage{array}
\usepackage[lofdepth,lotdepth]{subfig}

\usepackage{graphicx}
\usepackage{xcolor}
\usepackage{soul}

\usepackage{mdwmath}
\usepackage{amssymb}
\usepackage{cases}
\usepackage{tcolorbox}
\usepackage{authblk}
\begin{document}
 
%


\title{Reversible Privacy Preservation using Multi-level  Encryption and Compressive Sensing }

\author[1]{Mehmet Yama\c{c}}
\author[1]{Mete Ahishali}
\author[1]{Nikolaos Passalis}
\author[1]{Jenni Raitoharju}
\author[2]{Bulent Sankur}
\author[1]{Moncef Gabbouj}
\affil[1]{Tampere University, Faculty of Information Technology and Communication Sciences, Tampere, Finland}
\affil[2]{Bo\v{g}azi\c{c}i University, Electrical and Electronics Engineering Department, Istanbul, Turkey}

\maketitle

\begin{abstract}
Security monitoring via ubiquitous cameras and their more extended in intelligent buildings stand to gain from advances in signal processing and machine learning. While these innovative and ground-breaking applications can be considered as a boon, at the same time they raise significant privacy concerns. In fact, recent GDPR (General Data Protection Regulation) legislation has highlighted and become an incentive for privacy-preserving solutions. Typical privacy-preserving video monitoring schemes address these concerns by either anonymizing the sensitive data. However, these approaches suffer from some limitations, since they are usually non-reversible, do not provide multiple levels of decryption and computationally costly. In this paper, we provide a novel privacy-preserving method, which is reversible, supports de-identification at multiple privacy levels, and can efficiently perform data acquisition, encryption and data hiding by combining multi-level encryption with compressive sensing. The effectiveness of the proposed approach in protecting the identity of the users has been validated using the goodness of reconstruction quality and strong anonymization of the faces.
\end{abstract}



%
\IEEEpeerreviewmaketitle

\section{Introduction}

Modern intelligent buildings rely on efficient automation of various tasks ranging from traditional heating, ventilation, and air conditioning (HVAC) systems to advanced intelligent access control and monitoring systems, improving the quality of indoor  environment, for example, by ensuring a higher degree of safety by  hazard monitoring and by providing significant energy savings~\cite{weng2012buildings, nguyen2013energy}. However, accomplishing these tasks necessitate the installation of myriad sensors such as monitoring cameras, with concomitant privacy concerns for people living and/or working in such buildings. The recent General Data Protection Regulation (GDPR)~\cite{voigt2017eu} legislation in Europe reflects these concerns and regulates the ways in which sensitive data should be collected and processed, advocating  data and purpose minimization principles, i.e., limiting the collected data and the information that can be inferred from them to the minimum required by the corresponding application.

These issues are currently tackled by privacy-preserving approaches that either attempt to produce an anonymized version of the original data~\cite{agrawal2011person, du2014garp}, namely by obfuscating the sensitive parts of the images such as faces or employ data analysis methods with strong theoretical privacy guarantees, such as differential privacy~\cite{abadi2016deep} or homomorphic encryption schemes~\cite{lagendijk2013encrypted}. However, both approaches suffer from two main handicaps, first, as they are usually non-reversible, discarding thus the original sensitive information, and second, they are costly in energy consumption and demand high-performance data processing. An ideal privacy-preserving data analysis method should a) enable reversing the de-identification for authorized users without a significant cost in terms of processing power and bandwidth, b) not degrade the non-sensitive aspects of the data so that both the semi-authorized persons can analyze the scene, e.g., for abnormal behaviour but without identifying the people via their face images and the fully authorized persons can analyze the de-obfuscated signal, c) be lightweight in energy and computational requirements.

\begin{figure*}[]
 \centering
  \includegraphics[width=0.77\textwidth]{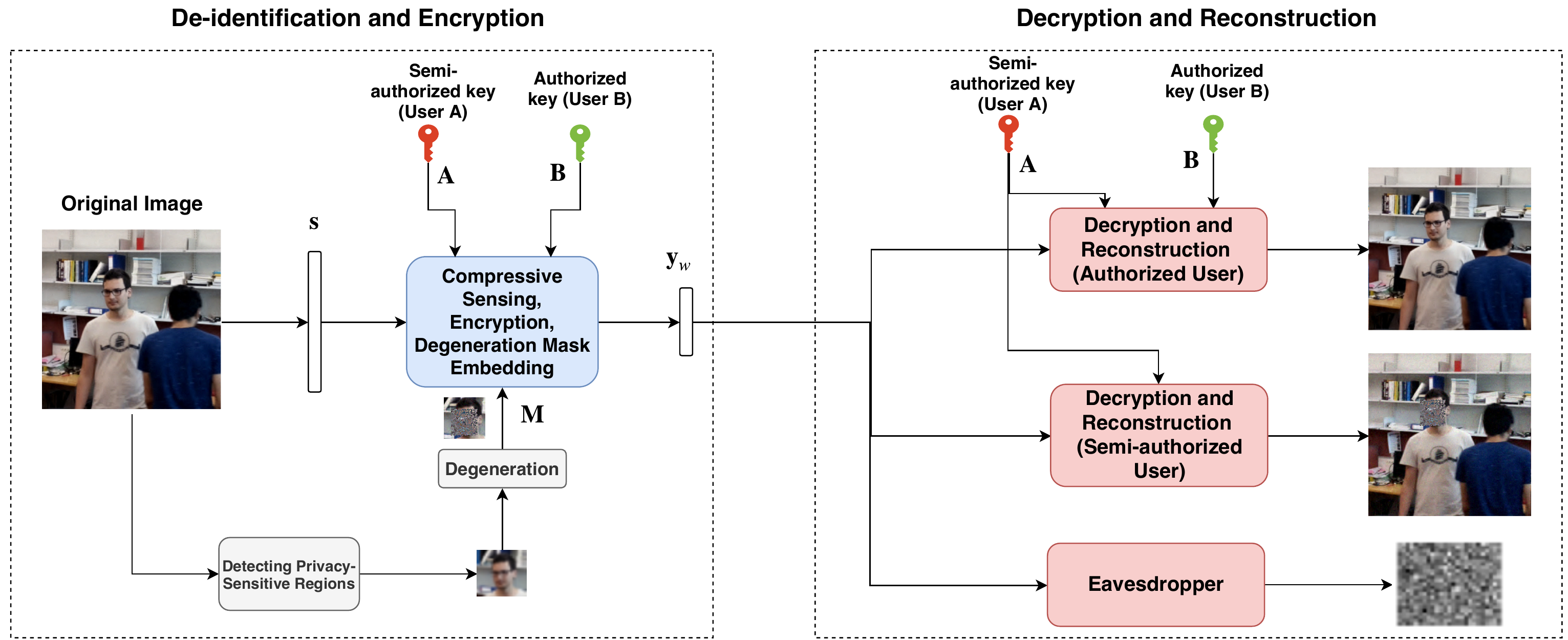}
\vspace{-0.2cm}
 \caption{Signal flow in the  proposed model and illustration of the three authorization levels.}
 \label{fig:model}
 \end{figure*}

In this work, we propose a novel and practical privacy-preserving solution for video monitoring applications such as in intelligent buildings and spaces under surveillance. Our method combines a multi-level encryption scheme with compressive sensing (CS). The proposed approach has two advantages over existing de-identification and privacy-preserving methods, namely a) it is reversible, that is, it is supporting de-identification at multiple privacy levels (as shown in Fig.~\ref{fig:model}, where different end users can get different levels of anonymized data according to the shared key, and b) it is resource and energy efficient, since it can jointly perform data acquisition, encryption, and transmission. Note that the proposed joint acquisition, privacy-protection and encryption scheme can be applied with few modifications to any kind of non-video, privacy-sensitive data.
 
In our method, the privacy-sensitive parts such as faces or other clues that will enable person identification for unauthorized or semi-authorized users, are obfuscated using a random corruption matrix, and then the compressed and encrypted signal is transmitted. We emphasize that compression and encryption are realized in one step via compressive sampling. The obfuscation matrix is separately encrypted and embedded into the transmitted signal. Note that the proposed method is capable of supporting multi-level de-identification, where a different level of recovery quality is provided for users at different authorization levels. The encryption,  compression and embedding of the obfuscation mask into the data stream are performed in a linear fashion allowing for fast and energy efficient implementations.  

\section{Background and Prior Work}

CS has greatly impacted various fields of signal processing since its inception in 2005 \cite{CS}. According to the CS theory, a signal can be sampled using far fewer measurements than Nyquist-Shannon type acquisition methods.  For instance, CS-based MRI imaging \cite{MR} or radar monitoring systems \cite{radar1, radar2} have been able to significantly reduce the signal acquisition time and bandwidth requirements over traditional approaches.  In addition to data acquisition advantage, CS is inherently cryptographic \cite{CSinEnc,CSinEnc2} since CS signals are linearly sampled using random  measurement matrices. CS-based encryption runs thus in parallel with the data acquisition and is a low-cost solution compared to well-known complex encryption standards such as AES or RSA.

\subsection{Compressive Sensing}
\label{Compressive Sensing Review}
Let $\mathbf{s} \in \mathbb{R}^N$ be an $N$-dimensional signal uniformly sampled using the traditional data acquisition scheme. In a  compressive sensing scheme, this signal is linearly sampled using $m << N$ measurements, i.e.,
\begin{equation}
  \mathbf{y} = \mathbf{A} \mathbf{s}, \label{eq_1}
 \end{equation}
where the matrix $\mathbf{A} \in \mathbb{R}^{m\times N}$ is denoted as the CS measurement matrix. It is well known that most of the signals we encounter in real life applications have some intrinsic  structure, allowing for sparse sampling in some proper domain, i.e., $\mathbf{s} = \bm{\Phi} \mathbf{x}$, where $\bm{\Phi}$ is generally an $N \times N$ sparsifying basis and $\mathbf{x}$ contains the sparse coefficients with $\left \| \mathbf{x} \right \|_0 \leq k$. The underdetermined system of equations in Eq.~\eqref{eq_1} can be uniquely solved under the sparsity constraint.  Using the knowledge of the sparsity or compressibility of signal decomposition in a proper domain (as given by basis $\bm{\Phi}$), we obtain the sparsest solution as:
\begin{equation}
    \hat{\mathbf{x}}=  \arg \min_\mathbf{x} \left \| \mathbf{x} \right \|_0 ~~ \text{s.t.} ~~ \left \| \mathbf{y} - \mathbf{H} \mathbf{x} \right \|_2 \leq \epsilon, \label{l0}
\end{equation}
where $\mathbf{H}=\mathbf{A} \bm{\Phi}$ and $\epsilon$ is the employed threshold for solving the corresponding problem. 

The optimization problem in Eq.~\eqref{l0} is non-convex and can be relaxed into the following $\ell_1$ minimization problem:
\begin{equation}
    \hat{\mathbf{x}}=  \arg \min_\mathbf{x} \left \| \mathbf{x} \right \|_1 ~~ \text{s.t} ~~ \left \| \mathbf{y} - \mathbf{H} \mathbf{x} \right \|_2 \leq \epsilon. \label{l1}
\end{equation} 
The uniqueness and stability conditions of Eq.~\eqref{l1} are well studied in the literature. The solution of this $\ell_1$ minimization problem is equal to the solution of the original $\ell_0$ minimization problem in Eq.~\eqref{l0} if the measurement matrix $A$ satisfies some conditions \cite{NullSpace,ca08} at the cost of an increase in the number of measurements. 


\subsection{One-class and Multi-class Encryption via Compressive Sensing}

The idea of using CS in a cryptosystem was first studied in \cite{Rachlin}, where Rachlin and Baron used the CS measurement matrix as an encryption key and investigated the possibility of reconstructing the sparse coefficients  $\mathbf{x}$ without the knowledge of $\mathbf{A}$. They have argued that even though it is not possible to achieve Shannon-perfect secrecy \cite{Shannon}, it is nevertheless unfeasible for an attacker to recover information in polynomial time~\cite{Rachlin}.  Then, it was proved that when the sampling matrix is i.i.d. Gaussian, then one can glean from measurement vector $\mathbf{y}$ at most information about the energy of the original signal~\cite{Bianchi}, thus achieving perfect secrecy (given that, $y$ is normalized to have a fixed energy in each transmission, the key $\mathbf{A}$ is sent in a secure channel and kept private). The robustness of the CS-based encryption was further studied in \cite{Orsdemir}, where it was pointed out that even if an adversary cannot estimate the secret key $\mathbf{A}$, he/she may try to ruin the communication with additive noise attacks. However, for CS-based cryptosystems, the overall system's robustness can be satisfied to a  degree even higher than that of traditional encryption schemes, provided certain necessary condition are met~\cite{Orsdemir}.

 A multi-class encryption scheme was also proposed in~\cite{Multiclass}, where the authors deliberately perturb the measurement matrix to adjust the recovery quality for different types of users by using a different measurement matrix $\mathbf{A} + \Delta \mathbf{A}$, where $\Delta \mathbf{A}$ is a partial perturbation matrix.  However, their application to reversible de-identification schemes is not straightforward since the corruption induced during the de-identification stage needs to be separately transmitted in a side channel. To this effect, the work in \cite{Yamac1, Yamac2} introduces such a steganographic channel that enables embedding some extra information directly on compressively sensed measurements. This steganographic channel and the resulting data hiding can allow efficient transmission of the de-identification information. Thus one can transmit encrypted and selectively obfuscated (anonymized) images with the frame-specific corruption matrix, which allows for reversing the de-identification process. Additionally, this approach further ensures the security of the system, since it is even harder for an adversary to decode the cover signal itself, given that an extra noise-like signal is added to the encrypted/compressed signal.

\section{Proposed Method}


We adopt the idea of using a perturbation matrix to obfuscate selectively the measurements related to the sensitive parts of images for privacy preservation. We propose a privacy-preserving method for acquiring, compressing and encrypting the information, but with the ability to reverse the de-identification so that an authorized person can recover the degraded part of the image using her key. This will be achieved without using an additional secure side channel to transmit the degradation mask.  The degradation mask can be transmitted within the compressed measurements stream. On the receiver end, three levels of security are conceived. An eavesdropper not knowing the CS measurement matrix $\mathbf{A}$, (called encryption matrix in the sequel) cannot decrypt the signal. A semi-authorized person (User A) knowing $\mathbf{A}$ can decrypt the signal, while she cannot recover the privacy-preserved, i.e., the degraded parts. Finally, an authorized person (User B), who possesses the key $\mathbf{k_b}$ can extract the degradation matrix and thus, recover the full image after having reconstructed the degraded part of the signal. The proposed method is illustrated in Fig.~\ref{fig:model}.

\subsection{Partial Perturbation of Encryption Matrix and Embedding} 

To conceal parts of image frames, we convert them into vectors, $\mathbf{s} \in \mathbb{R}^N$ and then define the vector indices, $j$, containing the privacy-sensitive sectors (this can be done manually or automatically, e.g., using face detection algorithms) and forming the set $\mathcal{C}$.
We degrade the encryption matrix, $\mathbf{A}$, as
\begin{equation}
    \tilde{\mathbf{A}} = \mathbf{A} + \mathbf{M},
\end{equation}
where $\mathbf{M} \in \mathbb{R}^{m\times N}$ is defined as:
\small
\begin{subnumcases}{{m}_{i,j}=}
   0  \hfill \text{ with probability $p$},   \hfill & \text{ if $ j \in \mathcal{C}$ } \\
   -2*A_{i,j} \hfill \text{ with probability $1-p$,}    & \text{ if $ j \in \mathcal{C}$ }
 \label{positive-subnum}
   \\
   0, & \text{ else. }  \label{negative-subnum}
\end{subnumcases}
\normalsize
Finally, we can encode the signal $\mathbf{s}$ with the degraded encryption matrix, $\tilde{\mathbf{A}}$, instead of $\mathbf{A}$, before broadcasting it, i.e., $\mathbf{y}_d = \tilde{\mathbf{A}} \mathbf{s}$. 

 

The perturbation matrix, $\mathbf{M}$, can be then ternary coded into a vector $\mathbf{w} \in \mathbb{R}^T$ as:
\small
\begin{subnumcases}{w_k=}
   a,    \hfill & \text{ if $m_{i,\mathcal{C}_k} = 0$ (for any row $i$),} \\
   -a, \hfill & \text{ if $ m_{i, C_k} = - 2* A_{i, \mathcal{C}_k}$ (for any row $i$), }
 \label{positive-subnum}
   \\
   0 \hfill &  \text{else,}  \label{negative-subnum}
\end{subnumcases}
\normalsize
where $\mathcal{C}_k$ is the $k$-th element of the set $\mathcal{C}$ and $a$ is the selected embedding power. Note that for a system that has a total data hiding capacity of size $T$, $|w| \le T$. 
The steganographic capacity $T$ is dictated by the data hiding limits~\cite{Yamac1, Yamac2}. At this stage, an embedding matrix (authorization key) $\mathbf{B} \in \mathbb{R}^{m \times T}, T < m$ is used to linearly embed the information for de-identification of the privacy-sensitive measurements directly into the encrypted signal:
\begin{equation}
    \mathbf{y}_w = (\mathbf{A} + \mathbf{M})\mathbf{s} + \mathbf{B}\mathbf{w} = \mathbf{H}\mathbf{x} + \mathbf{B}\mathbf{w} + \mathbf{n}
\end{equation}
where $\mathbf{x} \in \mathbb{R}^N$ is the vector of sparse coefficients of $\mathbf{s}$ in $\bm{\Phi}$, while the rest of the quantities are defined as $\mathbf{H} = \mathbf{A} \bm{\Phi}$, and $\mathbf{n} = M\mathbf{s}$.  We also impose an embedding power constraint $\left \| \mathbf{B}\mathbf{w} \right \| \leq P_E$ in order not to harm the reconstruction quality for the semi-authorized users. Finally, the encrypted measurements $\mathbf{y}_w$ are transmitted/stored as shown in the Fig. 1.

\subsection{Recovery Algorithms for Different Type of Users}
A semi-authorized user (User A), who has only the key $\mathbf{A}$, can apply the $\ell_1$-decoding scheme, as in Algorithm 1, to reconstruct the de-identified version of the video. The privacy-sensitive parts of the images, e.g., faces, will remain unrecognizable to user A. For the authorized users (User B), we follow the recovery method proposed in~\cite{Yamac2}. 
To extract $\mathbf{w}$, we first construct an annihilator matrix $\mathbf{F} \in \mathbb{R}^{ P \times m}$, so that $\mathbf{F}\mathbf{B}=0$ and $ P=m-T$. We remove this embedded signal by applying $\mathbf{F}$ to $\mathbf{y}_w$: \begin{equation}
    \tilde{\mathbf{y}} = \mathbf{F} (\mathbf{H}\mathbf{x} + \mathbf{B}\mathbf{w} + \mathbf{n}) = \mathbf{F}\mathbf{H}\mathbf{x} + \mathbf{z},
\end{equation}
where $\mathbf{z} = \mathbf{F}\mathbf{n}$. Then, a pre-estimation of the sparse signal $\mathbf{x}$ can be calculated as:
\begin{equation}
    \tilde{\mathbf{x}} = \arg \min \left \| \mathbf{x} \right \|_1 ~ s.t.~ \left \| \tilde{\mathbf{y}} - \mathbf{F}\mathbf{H}\mathbf{x} \right \|_2  \leq \epsilon.
\end{equation}
Hereafter, a pre-estimation of $\mathbf{w}$ can be found using least squares:
\begin{equation}
    \mathbf{w}'' = (\mathbf{B}^T\mathbf{B})^{-1} \mathbf{B}^T (\mathbf{y}_w - \mathbf{H} \tilde{\mathbf{x}}).
\end{equation}
Following this step,  the $0$'s in the ternary message can be extracted using a simple hard thresholding to $\mathbf{w}''$, i.e., ${\mathbf{\tilde{{w}}} }= \mathbf{{{w}''}}\odot  {\mathbf{{1}_{\left | {{{w}''}_i}\right| >\eta }}} $ where $\odot $ denotes the element-wise multiplication operator between two vectors,
\begin{subnumcases}{ {1}_{\left | {{w}}_i''  \right | >\eta, i} =}
   1  \hfill,   \hfill & \text{ if $\left | {{w}}_i''  \right | >\eta $,} \\
   0 \hfill & \text{ else,}
 \label{positive-subnum}
   \label{negative-subnum}
\end{subnumcases}
and $\eta$ is the threshold value. Finally, an improved estimate of the  embedded information is obtained as:
\begin{equation}
    \hat{w}_i=a*\text{sgn}(\tilde{w}_i),
\end{equation}
and the sparse signal is recovered as:
\small
\begin{equation}
    \hat{\mathbf{x}} = \arg \min_{\mathbf{x}} ~ \left \| \mathbf{x} \right \|_{\ell_1^N} ~~ \text{s.t.} ~~ \left \| (\mathbf{y}- \mathbf{B}\hat{\mathbf{w}})- (\mathbf{A} + \mathbf{M}) \bm{\Phi} \mathbf{x} \right \|_{\ell_2^m} \leq \epsilon.
\end{equation}
\normalsize
The recovery algorithm for the fully authorized user is provided in Algorithm~\ref{alg:2}. The recovery guarantee conditions, as well as a robustness analysis of the proposed embedding and recovery algorithm are provided in~\cite{Yamac1,yamac3}.

\begin{algorithm}
   \caption{Reconstruction for semi-authorized user}	 			   \label{previous_alg}
 	\begin{algorithmic}
 	\small
     	\State  \textbf{Input:} $\mathbf{y}$, $\mathbf{A}$, $\bm{\Phi}$;
    \State     \textbf{Hyper-parameters:} $\epsilon$
    
 \State  \textbf{1.} Estimate $\hat{\mathbf{x}}$ :  $\tilde{\mathbf{x}}=\arg \min_\mathbf{x} \left\|\mathbf{x} \right \|_{1} ~~ \text{s.t.} ~~\left \|\mathbf{y}_w-\mathbf{H} \mathbf{x}\right \|_2 \leq \epsilon $
 
  \State  \textbf{2.} $\hat{\mathbf{s}} = \bm{\Phi} \hat{\mathbf{x}}$.  
  
  
  \State \textbf{Return:} $\hat{\mathbf{s}}$       
   \end{algorithmic}
   \label{alg:old}
 \end{algorithm}


\vspace{-0.5cm}
 \begin{algorithm}
   \caption{Reconstruction for full-authorized user}	 			   \label{previous_alg}
 	\begin{algorithmic}
 	\small
     	\State  \textbf{Input:} $\mathbf{y}$, $\mathbf{A}$, $\mathbf{B}$, $\bm{\Phi}$;
    \State     \textbf{Hyper-parameters:} $\epsilon$
 \State  \textbf{1.} Apply $\mathbf{F}$ to $\mathbf{y}$ : $\tilde{\mathbf{y}} = \mathbf{F}\mathbf{y}$ 
  \State  \textbf{2.} Estimate $\tilde{\mathbf{x}}$ :  $\tilde{\mathbf{x}}=\arg \min_\mathbf{x} \left\|\mathbf{x} \right \|_1 ~~ \text{s.t.} ~~\left \|\tilde{\mathbf{y}}-\mathbf{F}\mathbf{H}\mathbf{x}\right \|_2 \leq \epsilon $
         \State 
  \textbf{3.} Estimate $\mathbf{w}''$ : $\mathbf{w}''= (\mathbf{B}^{\text{T}}\mathbf{B})^{-1}\mathbf{B}^{\text{T}} (\mathbf{y}-\mathbf{H}\tilde{\mathbf{x}}) $
  \State
  \textbf{4a.} Thresholding $\mathbf{w''}$: $\mathbf{\tilde{w} } = \mathbf{{w}''} \odot \mathbf{1}_{\left | {{{w}}_i'' } \right | >\eta   } $
  
  \State \textbf{4b.} Forming $\mathbf{\hat{w}}$, where $\hat{w}_i=a*\text{sgn}(\tilde{w}_i) $
  
  \State  \textbf{5.} Obtain $\mathbf{M}$ from $\hat{\mathbf{w}}$. 
  
  \State  \textbf{6.} $\hat{\mathbf{x}} = \arg \min_{\mathbf{x}} ~ \left \| \mathbf{x} \right \|_1 ~~ \text{s.t.} ~~ \left \| (\mathbf{y}- \mathbf{B}\hat{\mathbf{w}})- (\mathbf{A} +\mathbf{M}) \bm{\Phi} \mathbf{x} \right \|_2 \leq \epsilon$
  
  \State \textbf{7.} $\hat{\mathbf{s}} = \Phi \hat{\mathbf{x}}$.  
  
  \State \textbf{Return:} $\hat{\mathbf{s}}$       
   \end{algorithmic}
   \label{alg:2}
 \end{algorithm}

\subsection{Reversible Privacy-Preserving Video Monitoring}

Random, e.g., Gaussian measurement matrices are proven to be optimal for reconstruction performances. However, direct use of the proposed approach for compressing, encrypting and transmitting video frames is practically not feasible.  For example, consider a $512 \times 512$ image and  a measurement rate $\frac{m}{N} = 0.36$, i.e.,  90.000 measurements each demanding a different 512x512 random array, resulting in 80 gigabytes of memory. A feasible alternative for the measurement/encryption matrix consists of a matrix whose rows are randomly chosen from noiselet transform bases which are then permuted. Similarly, we pick the embedding matrix $B$ from the DCT transform rows in the same manner as these noiselet and DCT transforms. The sparsifying basis $\Phi$ is chosen as 2-D wavelet (DWT) transform. Note that these two bases choices, namely, DWT and noiselet, are based on them being as incoherent as possible. Furthermore,  recent studies \cite{preserving1, preserving2} also show that this type of CS matrix construction, i.e., choosing the rows of the matrix from a subset of the rows of a basis such as fractional Fourier, Hadamard, etc., still provides security guarantees.

\section{Experimental Evaluation}
\label{section:experiments}
The proposed reversible privacy-preserving method is evaluated on a typical video monitoring setup that is commonly employed in such applications as analytics for intelligent buildings, video surveillance, and intelligent access control systems. We have used two video sequences (10 minutes long, a total of 3,219 frames), which were captured from two different cameras installed at Tampere University in a  typical area monitoring setting.  The cameras provided two different views of an office environment.

\begin{figure}[]
 \centering
  \includegraphics[width=0.7\linewidth]{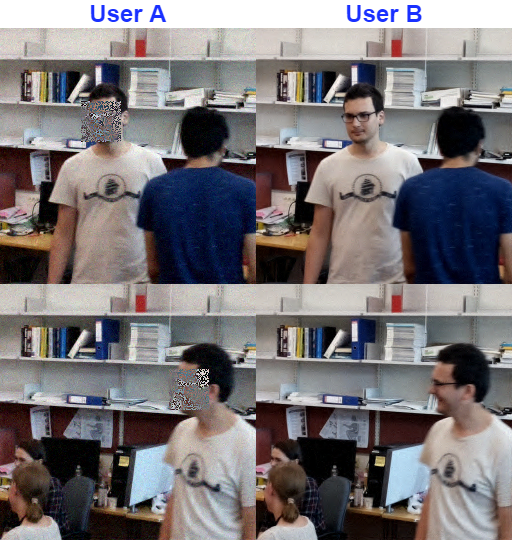}
  
\vspace{-0.2cm}
 \caption{Sample recovered frames for the semi-authorized (User A) and authorized (User B) (measurement rate 0.6)}
\label{fig:sample}
\end{figure}

First, the ability of the proposed method to compressively sense the scene while anonymizing the detected faces and reversing the de-identification process is demonstrated in Table~\ref{table:psnr}. For the conducted experiments we used $T=10000$ and $a$ is adjusted so that we can fix the embedding power to a predetermined small value i.e., $\frac{\left \| Bw \right \|_2}{ \left \| Ax \right \|_2} = 0.085$. The table reports the peak signal-to-noise ratios (PSNRs) averaged over 3,219 frames for anonymized and for clear parts of the transmitted frames. Notice that the semi-authorized user (A) receives very poorly reconstructed, practically unrecognizable face regions while the rest of the scene is well-reconstructed, only 3-5 dB worse than that of the authorized user (B).  For user B, both sensitive and non-sensitive parts of the frames are well reconstructed, the sensitive parts being slightly of low quality.  Two sample frames, where the information was decoded using the semi-authorized (User A) and the authorized (User B) key are shown in Fig.~\ref{fig:sample}

\begin{table}[h!]
\caption{Peak signal-to-noise ratios (PSNRs, dB)  for anonymized and clear parts of the video frames at various measurement rates (MRs)}
\vspace{-0.2cm}
\label{table:psnr}
\centering
\begin{tabular}{@{}c|cc|cc|cc@{}}
\toprule
\multicolumn{1}{c}{} & \multicolumn{2}{c}{\textbf{\begin{tabular}[c]{@{}c@{}} Concealed\\   Region\end{tabular}}} & \multicolumn{2}{c}{\textbf{\begin{tabular}[c]{@{}c@{}}Outside  of \\ Concealed Region\end{tabular}}} & \multicolumn{2}{c}{\textbf{\begin{tabular}[c]{@{}c@{}}Whole\\   Frame\end{tabular}}} \\ \midrule
\textbf{MRs} & \textbf{User A} & \textbf{User B} & \textbf{User A} & \textbf{User B} & \textbf{User A} & \textbf{User B} \\
0.3 & 10.23 & 21.59 & 24.27 & 26.19 & 21.79 & 25.81 \\
0.4 & 9.83 & 26.99 & 26.34 & 29.72 & 22.70 & 29.43 \\
0.5 & 9.57 & 31.30 & 28.06 & 33.16 & 23.26 & 32.90 \\
0.6 & 9.39 & 34.93 & 29.45 & 36.52 & 23.60 & 36.29 \\
0.7 & 9.27 & 38.19 & 30.57 & 39.68 & 23.81 & 39.50 \\
0.8 & 9.18 & 40.97 & 31.44 & 42.43 & 23.93 & 42.27 \\ \bottomrule
\end{tabular}
\end{table}
We also evaluated the ability of the privacy-preserving compression method to withstand automatic recognition attacks, where a deep learning algorithm was employed to recognize the persons that appeared in the frames. To this end, we employed a state-of-the-art pre-trained Convolutional Neural Network (CNN), as provided by the dlib library~\cite{dlib}. The used network achieves over 99\% recognition accuracy on standard Labeled Faces in the Wild benchmark dataset. For the recognition experiments, we used the pre-trained CNN to extract 128-dimensional face embeddings and query a database of known faces. Nearest neighbor search, considering the first nearest identity was used for the classification. In all the experiments 6 different identities were used. A total of 240 face images were collected and annotated (20 frames per identity were used to form the database, while the rest of them were used for testing). The experimental results are reported in Table~\ref{table:recongition}. The very low recognition rate is achieved for the semi-authorized user (close to random guessing, 16.67\%), while the same recognition rate as with the original frame is achieved when the measurement rate is higher than 0.6 (88.33\%). Therefore, the proposed method is capable of effectively protecting the privacy of the users against a state-of-the-art recognition method when the semi-authorized key was used while allowing the authorized users to recover the sensitive information.

\begin{table}[h]
    \caption{Face recognition accuracy (\%) when the semi-authorized (User A) and authorized (User B) key is used. Results for different measurement rates (MRs) are reported.}
    \vspace{-0.2cm}
    \label{table:recongition}
      \centering    
    \begin{tabular}{l|cccccc}
        \toprule
        \textbf{MR}& 0.3 & 0.4 & 0.5 & 0.6 & 0.7 & 0.8  \\
         \midrule
        \textbf{User A} & 23.33 & 18.33 & 22.50 & 6.67 & 25.00 & 15.83 \\ 
        \textbf{User B} & 44.17 & 73.33 & 81.67 & 86.67 & 89.17 & 88.33\\
        \bottomrule
    \end{tabular}

\end{table}

\section{Conclusions}

In this paper, we introduced a novel privacy-preserving compression method, which is reversible and supports de-identification at multiple privacy levels. The method can be efficiently implemented to jointly perform data acquisition, encryption, and transmission by combining multi-level encryption with compressive sensing. We implicitly assumed the availability of an ancillary technique to identify and select the privacy-sensitive part of a signal; for example, a face detector in video surveillance. The viability of the proposed approach in protecting the identity of the faces was validated using both PSNR reconstruction measure, as well as by demonstrating its robustness against automated machine learning attacks.  
Future work will proceed for the sequential update of the anonymized regions. More extensive testing will be carried over system parameters using open surveillance video databases. The degree of obfuscation needs to be tested also with human evaluators.

\section*{Acknowledgements}
This work was supported by a NSF-Business Finland CVDI project (Amalia 3333/31/2018) and a Business Finland project VIRPA D (7940/31/2017) sponsored by Tieto Oyj, CA Technologies, and other companies.





%
\bibliographystyle{IEEEtran}
\bibliography{references}

\end{document}